\documentclass[12pt,titlepage,manuscript]{article}

 \usepackage{psfig}
\begin{document}
\noindent{\Large \bf
Emerging Magnetism in Platinum Nanowires 
}
\\
\\
{\small
A. Delin$^1$ and E. Tosatti$^{1,2,3}$ \\
$^1$Abdus Salam International Center for Theoretical Physics (ICTP), Strada Costiera 11, 34100 Trieste, Italy\\
$^2$International School for Advanced Studies (SISSA), via Beirut 2--4, 34014 Trieste, Italy\\
$^3$INFM DEMOCRITOS National Simulation Center, via Beirut 2--4, 34014 Trieste, Italy
			              \\
				      \\
}
\\
\\
{\bf Abstract} \\
We have investigated infinitely long, monostrand Pt nanowires theoretically, and found that
they exhibit Hund's rule magnetism. We find a  spin moment of 0.6~$\mu_B$ per atom, at the equilibrium bond length. 
Its magnetic moment increases with stretching.
The origin of the wire magnetism is analyzed and its effect on the conductance through the wire is discussed.

\newpage
\section{Introduction}
Little is currently understood about how
magnetism arises at the nanoscale and how it affects the properties of nanosized metallic objects.  
Systems of special interest in this context are nanowires and atomic-sized 
nanocontacts. The one-dimensional (1D) character of 
such systems causes specific physical phenomena to appear, 
most notably quantized ballistic conductance.\cite{wees1988}

These phenomena will be heavily affected by the possible
presence of magnetism in the nanosystem, especially of a 
genuine Hund's rule magnetic order parameter. 
Here, we report theoretical studies of magnetism in Pt 
monowires, i.e., wires consisting of a single line of equally spaced Pt atoms. 
Such objects, several atoms long, have recently been observed by Rodrigues {\it et al.}\cite{rodrigues}

Pt is a transition metal, with a partially filled $5d$ shell. 
The free Pt atom has a $d^9s^1$ configuration, giving a total magnetic moment of 2~$\mu_B$, and therefore it is 
reasonable to expect that 
a sufficiently stretched wire should eventually show some type of magnetic ordering. In particular,
at the localized side of a Mott transition, one could expect an antiferromagnetic ordering.
In the bulk metal, the Pt $5d$ band is too wide to provoke spin-polarization and  
there is no evidence of Pt showing surface magnetism either.\cite{blugel1992}
Yet, the $5d$ band is still only partially occupied, and the density of states at the Fermi level is quite high.
It appears therefore to be an open question whether Pt nanowires might exhibit ferromagnetic ordering.
If that were the case, it might result in interesting phenomena such as spin-polarized current flow and magnetic-field dependent conductance,
phenomena that could prove highly useful in spintronics applications.

Of course, thermal fluctuations, very large in a nanosystem, will generally 
act to destroy static magnetic order in the absence of an 
external field. 
Sufficiently slow fluctuations transform a nanomagnet to a superparamagnetic 
state, where magnetization fluctuates on some time scale,
between equivalent magnetic 
valleys, separated, e.g., by anisotropy-induced energy 
barriers. If the barriers are sufficiently large 
--- and the fluctuations sufficiently slow --- the nanosystem spends most 
of the time in a single magnetic valley, and will for many 
practical purposes behave as magnetic. We may in these circumstances
be allowed to neglect fluctuations altogether, and to 
approximate the calculated properties of the 
superparamagnetic nanosystem with those of a statically magnetized one.
Experimentally, evidence of 1D superparamagnetism  with fluctuations
sufficiently slow on the time scale of the probe has been reported
in Co atomic chains deposited at Pt surface steps.\cite{gambardella2002}

\section{Method}
The density-functional calculations\cite{dft} reported here
are all-electron, in order to rule out possible sources of doubt that might arise when using
pseudopotentials in presence of magnetism and in a nonstandard
configurations.\cite{bahn2001} 
We employed an all-electron full-potential linear
muffin-tin orbital (FP-LMTO) basis set\cite{wills} together with 
a generalized gradient approximation (GGA)\cite{gga} to the 
exchange-correlation functional. As a double check, some of the 
calculations were repeated using the linear augmented plane-wave 
(LAPW) code WIEN97.\cite{wien97} None of these calculations 
assume any shape approximation of the potential or wave functions.

We performed both scalar relativistic (SR) calculations, and calculations including the spin-orbit
coupling as well as the scalar-relativistic terms. 
The latter will be referred to as ``fully relativistic'' (FR) calculations in the following,
although we are not strictly solving the full Dirac equation, or making use of current density functional theory.
In the FR calculations, the spin axis was chosen to be 
aligned along the wire direction.

The calculations were performed with inherently three-dimensional
codes. Thus, the infinitely long, straight, isolated monatomic nanowire 
was simulated by a regular array of well-spaced nanowires. 
Convergence of the magnetic moment was checked with respect to k-point mesh density, Fourier mesh density,
tail energies, and wire-wire vacuum distance.

\section{Results and Discussion}

\begin{figure}[h]
 \psfig{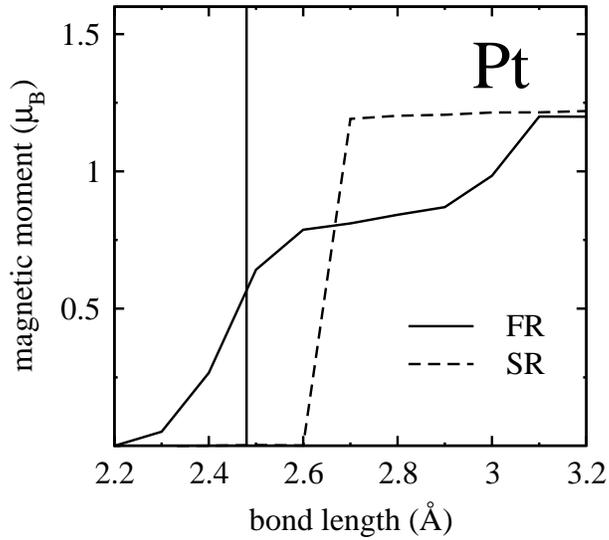}
 \caption
  {
  Magnetic spin moment per atom as a function of bond length for a long, monatomic wire of Pt.
  The vertical line points out the equilibrium bond length. 
  (FR = fully relativistic calculation;  SR = scalar relativistic calculation).
  See also Ref.~\protect\cite{delin2003}.
 \label{fig:ptmw_magnetic_moment}
  }
 \end{figure}

The magnetic spin moment per atom monowire as a function of bond length is shown in Fig.~\ref{fig:ptmw_magnetic_moment}.
The solid line refers to the fully relativistic calculation (FR), and the dotted line to the scalar relativistic (SR)
calculation.
As seen, the magnetic profiles for the SR and FR calculations are very different.
The SR calculation for Pt predicts this metal to be magnetic only for wire bond lengths larger than around 2.7~{\AA}, 
which corresponds to a rather stretched wire, 
whereas the FR calculation predicts it to
be magnetic in the whole range of bond lengths plotted (2.2~{\AA} to 3.2~{\AA}), with a moment of 0.6~$\mu_B$ per atom at the
equilibrium bond length 2.48~{\AA}.
For sufficiently large bond lengths (in the interval shown in the figure)
the magnetic moment reaches a plateau value, still well below  the atomic spin moment 2~$\mu_B$.
For even larger bond lengths (not shown), the magnetic moment eventually 
approaches the atomic spin moment 2~$\mu_B$.
The energy gain per atom due to spin polarization is rather small, around 8\,meV
for the FR calculation at the equilibrium bond length 2.48~{\AA}.
Antiferromagnetic Pt monowire configurations were also tested for bond lengths around the equilibrium one, and found to be energetically unstable
compared to the ferromagnetic configuration.

Several mechanisms, some favoring and other disfavoring a spin-polarized ground state, are at work in the wire.
The number of nearest neighbors is only two in the wire, compared to 12 in bulk.
This reduction of the number of nearest neighbors in the wire compared to the bulk
causes a narrowing of the $5d$ band, and the band width may become 
sufficiently small that the gain in exchange energy due to spin polarization is larger than the increase in kinetic energy.
On the other hand, the equilibrium bond length is significantly smaller in the wire, 10\% smaller than in bulk, which partly counteracts the
band-narrowing effect of the reduced number of neighbors.

Highly important in this context are the very sharp van Hove singularities caused by the one-dimensionality of the system.
These give rise to a very high density of states at the Fermi level when a van Hove singularity is sufficiently close to the 
Fermi level, which in turn results in a Stoner product larger than one, and thus spontaneous spin-polarization.

\begin{figure}[h]
 \psfig{figure=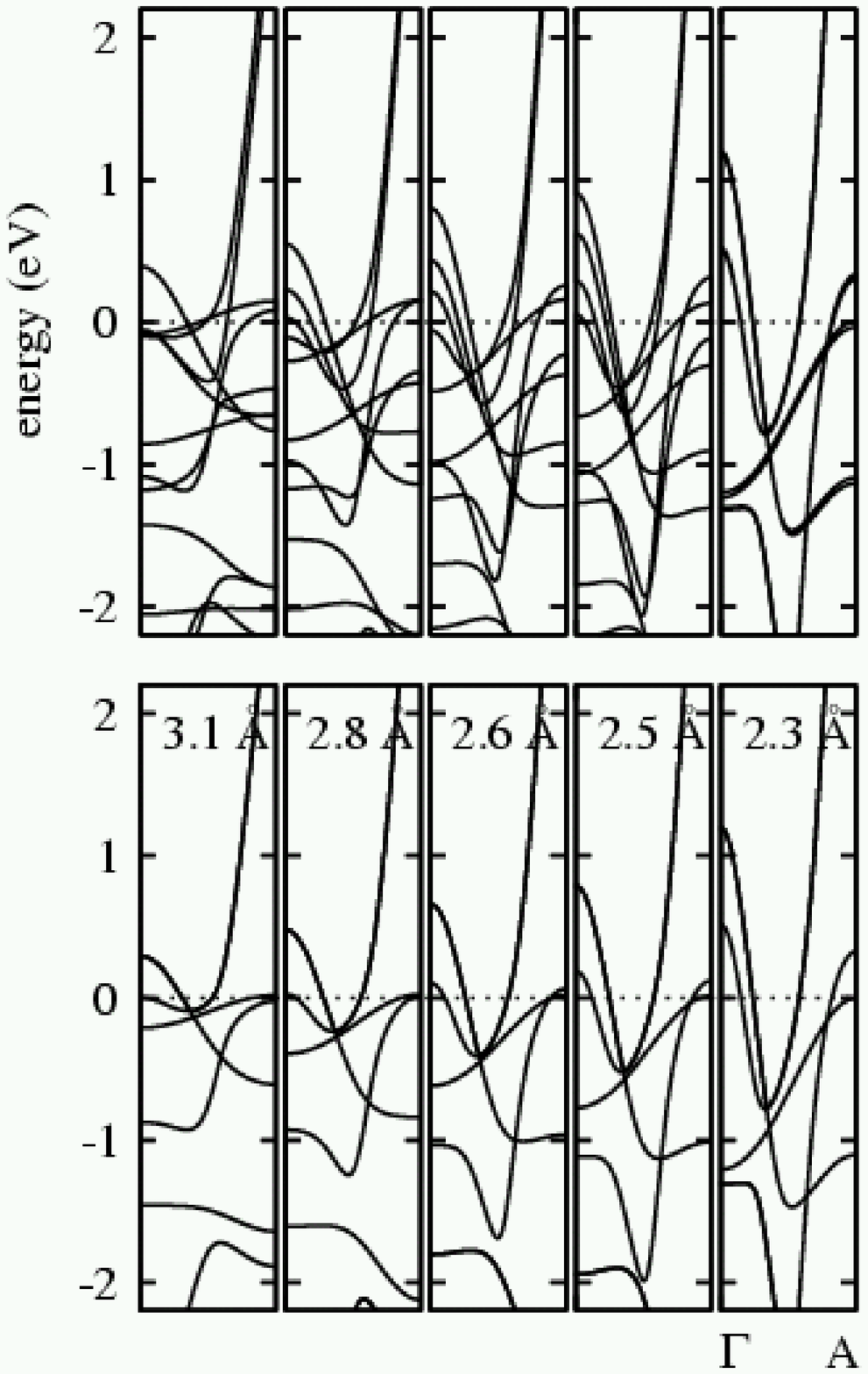,width=8.0cm}
 \caption
  {
  Fully relativistic Pt wire band structures, along the wire direction,
  for the five bond lengths 3.1~{\AA}, 2.8~{\AA}, 2.6~{\AA}, 2.5~{\AA} and 2.3~{\AA}.
  The Fermi energy is at zero. 
  The upper panels show the ferromagnetic case, and the lower panels the 
  nonmagnetic case.
 \label{fig:ptmw_bandstructures_soc}
  }
 \end{figure}

\begin{figure}[h]
 \psfig{figure=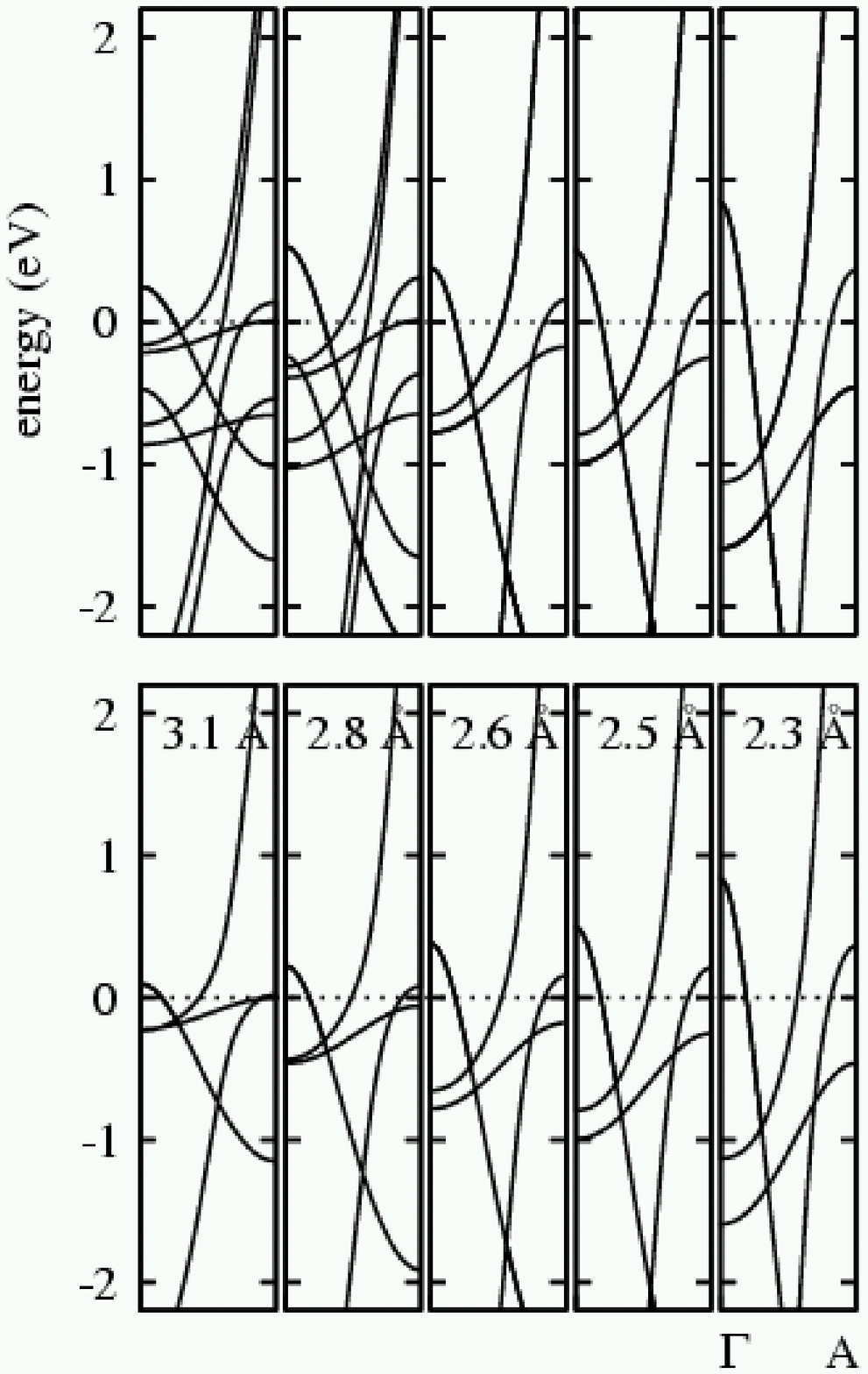,width=8.0cm}
 \caption
  {
  Scalar-relativistic band structures, along the wire direction,
  for the five bond lengths 3.1~{\AA}, 2.8~{\AA}, 2.6~{\AA}, 2.5~{\AA} and 2.3~{\AA}.
  The Fermi energy is at zero. 
  The upper panels show the ferromagnetic case, and the lower panels the 
  nonmagnetic case.
 \label{fig:ptmw_bandstructures_nosoc}
  }
 \end{figure}

\begin{figure}[h]
 \psfig{figure=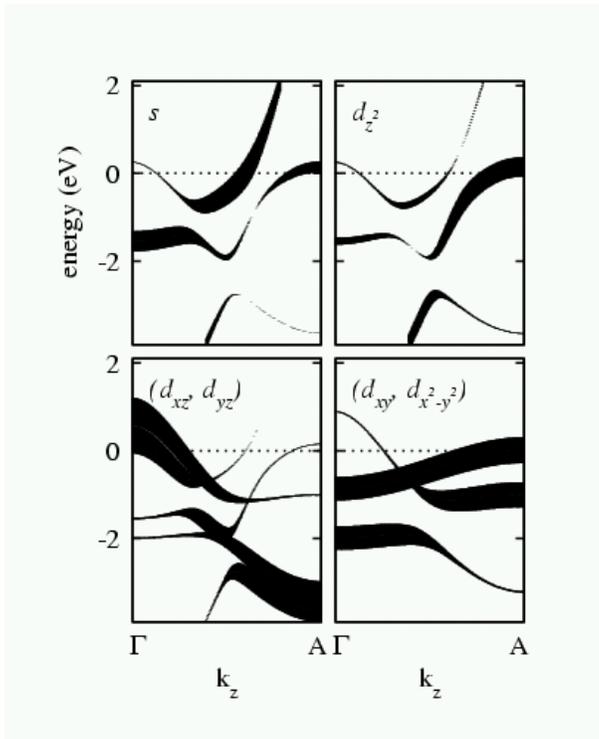,width=8.0cm}
 \caption
  {
  Character-resolved fully relativistic band structure along the wire direction, 
  for nonspinpolarized Pt with a bond length of 2.5~{\AA}.
  The Fermi energy is at zero.
  See also Ref.~\protect\cite{delin2003}.
 \label{fig:ptmw_fatbands}
  }
 \end{figure}

In order to provide a more detailed analysis of the origin of the magnetism in the wires, we found it useful to analyze 
the wire band structures.
Fig.~\ref{fig:ptmw_bandstructures_soc} shows FR band structures of the Pt monowire for several different bond lengths.
The bands run from the zone center, $\Gamma$, to the zone edge, A, in the direction of the wire.
Fig.~\ref{fig:ptmw_bandstructures_nosoc} shows the corresponding SR band structures.
The sharp van Hove singularities manifest themselves in the band structures 
through the horizontal band edges at the zone edge A and zone center $\Gamma$.
Since the orbital character is of critical importance for the moment formation, it is useful to divide the band structure into distinct orbitals ---
illustrated in Fig.~\ref{fig:ptmw_fatbands} for the FR case.
This figure has four panels, displaying separately the
$s$, $d_z$, $(d_{xz},d_{yz})$ and $(d_{xy},d_{x^2-y^2})$ characters
of the bands.
The vertical error bars, or ``thickness'', of the bands indicate the relative character weight.
In the SR case (not shown), the $(d_{xz},d_{yz})$ and $(d_{xy},d_{x^2-y^2})$ orbitals correspond each to separate bands, and the
$s$ and $d_z$ orbitals hybridize, forming two bands of high dispersion.
With the help of  Fig.~\ref{fig:ptmw_fatbands}, we can immediately recognize an important mechanism that favors spin-polarization in the wire.
As seen, the bands have mostly $d$ character at the edges, and therefore the exchange energy
gain will be rather large if a band spin-splits so that one of
the spin-channel band edges ends up above the Fermi level, and the other one below.
\footnote{Strictly speaking, the spin-orbit coupling will mix the two spin channels so that, in general, an eigenvalue
will have both majority and minority spin character.
However, we found that this mixing is so small, typically just a few percent, that it is
irrelevant for the qualitative discussion we make here.}
Thus, if a band edge ends up sufficiently near the Fermi level, we may expect a magnetic moment
to develop. While apparently similar to the magnetization of the jellium
wire,\cite{zabala1998} magnetism here is much more substantial, since the $d$ states involve
a much stronger Hund's rule exchange. 
We say that the magnetism exhibited in the Pt wire is Hund's rule magnetism, 
in order to differentiate it from the situation in the jellium wire.

By comparing the magnetic and nonmagnetic band structures, we see that 
the relatively flat $(d_{xy},d_{x^2-y^2})$ bands play the leading role in the formation of the magnetic state.
In the SR band structures,
these bands (all degenerate) sit well below the Fermi level, between $-1$~eV and $-0.2$~eV for the equilibrium bond length, 
and thus cannot contribute to spin polarization.  For larger bond lengths than around 2.6 {\AA}, the edge at A moves critically 
close to the Fermi level, and the bands split.
In the FR calculations, however, a band edge at A of this symmetry is pinned to the Fermi level 
in the whole range of studied bond lengths, down
to 2.2~{\AA}, explaining the large difference in magnetic profile between the SR and FR calculations. 
This band edge is close to the Fermi level in the FR calculation but not in the SR calculation simply because
in the FR calculation, the $(d_{xy},d_{x^2-y^2})$ orbital hybridizes partly with the $(d_{xz},d_{yz})$ orbital, with 
the result that the  $(d_{xy},d_{x^2-y^2})$ band splits into an upper part of purely  $(d_{xy},d_{x^2-y^2})$ character, 
and a lower part hybridizing strongly with the  $(d_{xy},d_{x^2-y^2})$ orbital, thereby splitting up into two bands.
If the average position of all the $(d_{xy},d_{x^2-y^2})$ levels is taken, we end up more or less where the SR  $(d_{xy},d_{x^2-y^2})$  orbital
originally was, energy-wise. Thus, 
one net effect of the spin-orbit coupling is that it increases the energy of part of the $(d_{xy},d_{x^2-y^2})$ orbital.
This also explains why the magnetic moment of the FR calculation is smaller than in the SR calculation in the
bond-length range 2.7~{\AA} to 3.1~{\AA}.

Three other band edges,
a $d_{z^2}$-dominated one located at A and the $(d_{xz},d_{yz})$-dominated ones
with located at $\Gamma$ are also important, and add to the size of the magnetic moment as they split around the 
Fermi level once the spin-polarization has been triggered.

The magnetic or superparamagnetic state of a small piece of 
nanowire bridging between nonmagnetic tips might at first seem 
very problematic to detect. It should in fact be detectable by
measuring ballistic conductance as a function of both temperature
and of an external magnetic field. The field, even a modest one 
depending on temperature, can drive a superparamagnetic wire 
from thermally disordered to fully spin polarized. In
this polarized state, the nature and number of current-carrying channels,
each corresponding to a band crossing the Fermi level, will
differ from that of the nonmagnetic state. 
The channel number characteristic of the magnetic state should in fact
last well into the superparamagnetic state at zero field. 
In the case of our infinitely long Pt monowire, we find that the nonmagnetic wire has 10 open conductance channels and
the spin-polarized 8, which would correspond to a Landauer conductance of $5 G_0$ and $4 G_0$, respectively,
if all channels conducted fully. ($G_0 = 2e^2/h$ is the fundamental conductance quantum.)
However, the bands splitting around the Fermi level due to the magnetic state
are of mainly $d$-character, which conduct poorly compared to $s$ channels. 
This means that the conductance of the magnetic and nonmagnetic wires will be much 
lower than these numbers indicate, and the difference in conductance will be significantly
smaller than $G_0$. Conductance calculations for a magnetic Pt nanowire segment between tips are
presently being considered, but are beyond the scope of this work.

In summary, we find that infinitely long Pt monowires have a ferromagnetic ground state, with a moment of
around 0.6~$\mu_B$ at the equilibrium bond length 2.48~{\AA}. The moment increases with stretching, and the 
trigger of the moment formation is the  $(d_{xy},d_{x^2-y^2})$ orbital.
The resulting superparamagnetic state of the nanowire will 
show up in the ballistic conductance in the form of a strong 
and unusual magnetic field and temperature
dependence. 
Also, more majority bands cross the Fermi level than do minority bands, 
resulting in a partial spin-polarization of the transmitted electron
current. If this current could be measured, it would be a very direct way of 
confirming the existence of a superparamagnetic state.
Rodrigues {\it et al.}\cite{rodrigues} recently measured the charge conductance of Pt nanocontacts and 
found features above as well as below $G_0$. More theory work will be needed to address their data, 
explicitly including such elements as tips and temperature.
\\
\\
{\bf Acknowledgments} \\
A.D. acknowledges financial support from
the European Commission through contract no. HPMF-CT-2000-00827 Marie Curie fellowship,
STINT (The Swedish Foundation for International Cooperation in Research and Higher Education),
and NFR (Naturvetenskapliga forskingsr{\aa}det).
Work at SISSA was also sponsored through TMR FULPROP, MUIR (COFIN and FIRB RBAU01LX5H) and by INFM/F.
Ruben Weht is acknowledged for discussions, and for double-checking some of the calculations using the WIEN97 code.
J. M. Wills is acknowledged for letting us use his FP-LMTO code.
We are also grateful to D. Ugarte for sharing with us the unpublished results of Ref.~\cite{rodrigues}.

%%%%%%%%%%%%%%% REFERENCES 

%
%%%%%%%%%%%%%%% FIGURES
% MAGNETIC MOMENTS
%
% BAND STRUCTURES  soc
%
% BAND STRUCTURES  nosoc
%
% FAT BANDS 

\end{document}